\documentstyle[11pt,paspconf,epsf,twoside,makeidx]{article}
\makeindex
\begin{document}

\title{Introduction to Principal Components Analysis\altaffilmark{1}}
\author{Paul J. Francis}
\affil{Department of Physics \& Theoretical Physics,
Australian National University, Canberra, Australia;
pfrancis@mso.anu.edu.au}
\author{Beverley J. Wills}
\affil{McDonald Observatory \& Astronomy Department, University of
Texas at Austin, TX, USA 78712; bev@astro.as.utexas.edu}
\altaffiltext{1}{This paper is for our friends Leah Cutter and Mike
Brotherton, who announced their engagement during the La Serena meeting.}
\begin{abstract}

Understanding the inverse equivalent width -- luminosity relationship
(Baldwin Effect), the topic of this meeting, requires extracting information 
on continuum and emission line parameters from samples of AGN.  We wish to 
discover whether, and how, different subsets of measured parameters may correlate
with each other.  This general problem is the domain of Principal Components
Analysis (PCA).
We discuss the purpose, principles, and the interpretation of PCA, using some
examples from QSO spectroscopy.   The hope is that identification of 
relationships among subsets of correlated variables may lead to new physical
insight.

\end{abstract}


\section{Introduction}

The point of all statistics is simplification. The amount of data the
world can throw at us would swamp Einstein: we have to simplify to
survive. Statistics is the art of extracting simple comprehensible
facts that tell us what we want to know for practical reasons, from
the floods of data washing over us.

Consider the fuel consumption of cars, for example. Every car will be
different, depending on its model, year, maintenance state, and the
aggression level of its driver. To fully characterize the fuel economy
of cars in the USA would require a different number for every car/driver
combination: that is, more than $10^8$ numbers. For most purposes, however,
such as working out the nation's likely oil usage, these $10^8$ numbers
can be replaced with one: the average fuel consumption. An enormous
simplification!

Principal Components Analysis (PCA) is a tool for simplifying one particular
class of data. Imagine that
you have $n$ objects (where $n$ is large), and you can measure $p$
parameters for each of them (where $p$ is also large). For example, the
objects could be the $n$ QSO
researchers attending a meeting in La Serena, and the parameters could
be the $p$ things you know about each of them: e.g., their heights, weights,
number of publications, 
\index{frequent flier miles}frequent flier miles and the fuel consumption of
their cars. Let us imagine that you want to investigate how these $p$
parameters are related to each other. For example, do astronomers who
spend most of their lives in airports publish more? Does this depend on
how fat they are? Do people with inefficient cars fly more, or is it
only the smart ones (with lots of publications) that do so? Do these
correlations represent real causal connections, or it it just that once 
you get \index{tenure}tenure you buy a 
\index{car!new}new car, become fat, stop publishing and give lots
of invited talks in exotic foreign locations?

The traditional way of dealing with this type of problem is to plot
everything against everything else and look for correlations. Unfortunately,
as the number of parameters increases, this rapidly becomes impossibly
complicated. It is easy to get lost in the web of parameters, each of
which correlates more or less well with some combination of the other
parameters. The human brain can cope with two or three parameters easily. By
plotting all the different variables against each other separately, we
can just about learn something about 5 -- 7 variables. But once we are
beyond this, the human brain needs help.

PCA is specifically designed to help in situations like this: when you
know lots of things about lots of objects, and want to see how all these
properties are inter-related. Basically, PCA looks for sets of parameters
that always correlate together. By grouping these into one new parameter,
an enormous saving in complexity can be achieved with minimal loss of
information.

PCA is one of a family of algorithms (known as multivariate
statistics) designed to handle complex problems of this sort. It
was first widely applied in the social sciences. The most infamous
early application of PCA was to intelligence testing. You can
test the intellectual ability of people in many ways. For example, you could
give a sample of $n$ people a set of $p$ exams, with questions testing their
creativity, memory, math skills, verbal skills etc. Do people who
score well on one test score well on all? Or do the scores break up into
sub-groups, such as verbal or logical scores, which correlate well with
the scores on other similar tests? PCA was applied to these exercises, and
it was found that nearly all the scores correlate well with each other.
Thus, it was claimed, a single underlying variable (known as IQ) can
be used to replace all the individual scores, and once you know someone's
IQ, you can accurately predict their performance on all the tests.
(See Steven Jay Gould's `The Mismeasure of Man' for a hilarious account
of the misuse of this application of PCA.)


\section{Overview}

\index{corr!eigen!defn}
The task of PCA is then, given a sample of $n$ objects with $p$ measured quantities
for each, i.e.  $p$ variables, $x_j$  ($j$ = 1, \ldots, $p$), find a set of $p$ new,
orthogonal (i.e. independent) variables, $\xi_1, \ldots , \xi_i, \ldots , \xi_p$, each
one a linear combination of the original variables, $x_j$:
$$ \xi_i = a_{i1} x_1 + \cdots + a_{ij} x_j + \cdots + a_{ip} x_p $$
Determine the constants $a_{ij}$ 
such that the smallest number of new variables account for as much of
the variance of the sample as possible.
The $\xi_i$ are called principal components.

If most of the variance
in the original data can be accounted for by just a few of the $p$ new
variables, we will have found a simpler description of the original dataset.
A smaller number of variables may point to a way of classifying the data.
More interesting, beyond the realm of statistical description, the PCA,
by showing which original variables correlate together, may lead to new
physical insight.
Of course it will sometimes happen that the observed variables are uncorrelated, 
or at least, lead to no dominant principal components.  That may be useful to
know, but not very interesting.

The concept of PCA is usually introduced either algebraically, through covariance
matrices, or geometrically.  We will first give a geometrical overview, then
illustrate with examples and interpretation.
Many textbooks on multivariate statistics give rigorous mathematical
treatments (e.g., Kendall 1980).

\section{A Geometrical Approach to Principal Components Analysis}
Consider the case of $p$ variables.  The data of $n$ QSOs are represented
by a large cloud in $p$-dimensional space.  If two or more parameters are correlated,
the cloud will be elongated along some direction in hyper-space defined by their
axes.  Large extensions can arise when a few parameters are correlated, or when
smaller correlated variations occur for a substantial number of variables.

\index{corr!eigen!defn}
PCA identifies these extended directions and uses them as a set of axes for the parameterization
of the multidimemsional space.  Following the analysis, each QSO can be represented by
its coordinates in the new space.
The new axes are identified sequentially: PCA first finds the most extended
direction in the original $p$-dimensional space by minimizing the sums of squares of
the deviations from that direction.  This direction forms the first principal
component (often called eigenvector 1), and accounts for the largest single linear
variation
among measured QSO properties.  Next we consider the ($p-1$)-dimensional hyper-plane
orthogonal to the first principal component.  We then search for the direction that
represents the greatest variance in ($p-1$)-space, thus defining the second principal
component.  This process is continued, defining a total of $p$ orthogonal directions.

\section{Examples using Real Data}
\subsection{PCA with Two Variables}

\index{corr!eigen!opt}
Consider the case of 22 QSOs each with measured values of X-ray spectral
index $\alpha_x$ (defined by F$_\nu \propto \nu^{-\alpha_x}$ between 0.15 keV and
2keV), and FWHM H$\beta$ (full width at half maximum for the broad H$\beta$
emission line).  The data points are distributed in an elongated cloud in 2
dimensions, as shown in Fig. 1.   It is standard practice to subtract the mean
value from each variable, and normalize by dividing by the standard deviation.
One can find the direction of the first principal component axis by rotating an
axis to align with  the direction of maximum elongation, actually maximum variance,
of the data.  The result of this is shown by the dashed line labeled PC1 in Fig. 2.
Because the points remain the same distance from the origin, by Pythagoras'
theorem, maximizing the variance along PC1 is equivalent to minimizing the sums
of squares of the distances of the points from this line through the origin;
these distances are 
shown as dotted lines.  The distance of a point from the origin, projected onto 
the direction PC1 represents the value (score) of the first principal component
for that data point.  Clearly, PC1 is a linear combination of the original input
variables.  The variance of PC1 is 1.764 -- rather more than the unit variance of
the original variables.  The total variance of the sample is the sum of the
variances for each variable, in this case, 2.  Because the new coordinates are
found simply by rotation, the distance of the points from the origin remains
unchanged, so the total variance remains the same.  Thus the first
principal component accounts for 1.764/2 = 88.2\% of the total sample variance. 
\index{corr!eigen!opt}
The remaining variance of the sample can be accounted for by the projection of the
data points onto the axis PC2, perpendicular to PC1 -- or 0.236/2 = 11.8\%.
These projections (lengths of the dotted lines) are the values or scores of the
second principal component.  

\begin{figure}
\plottwo{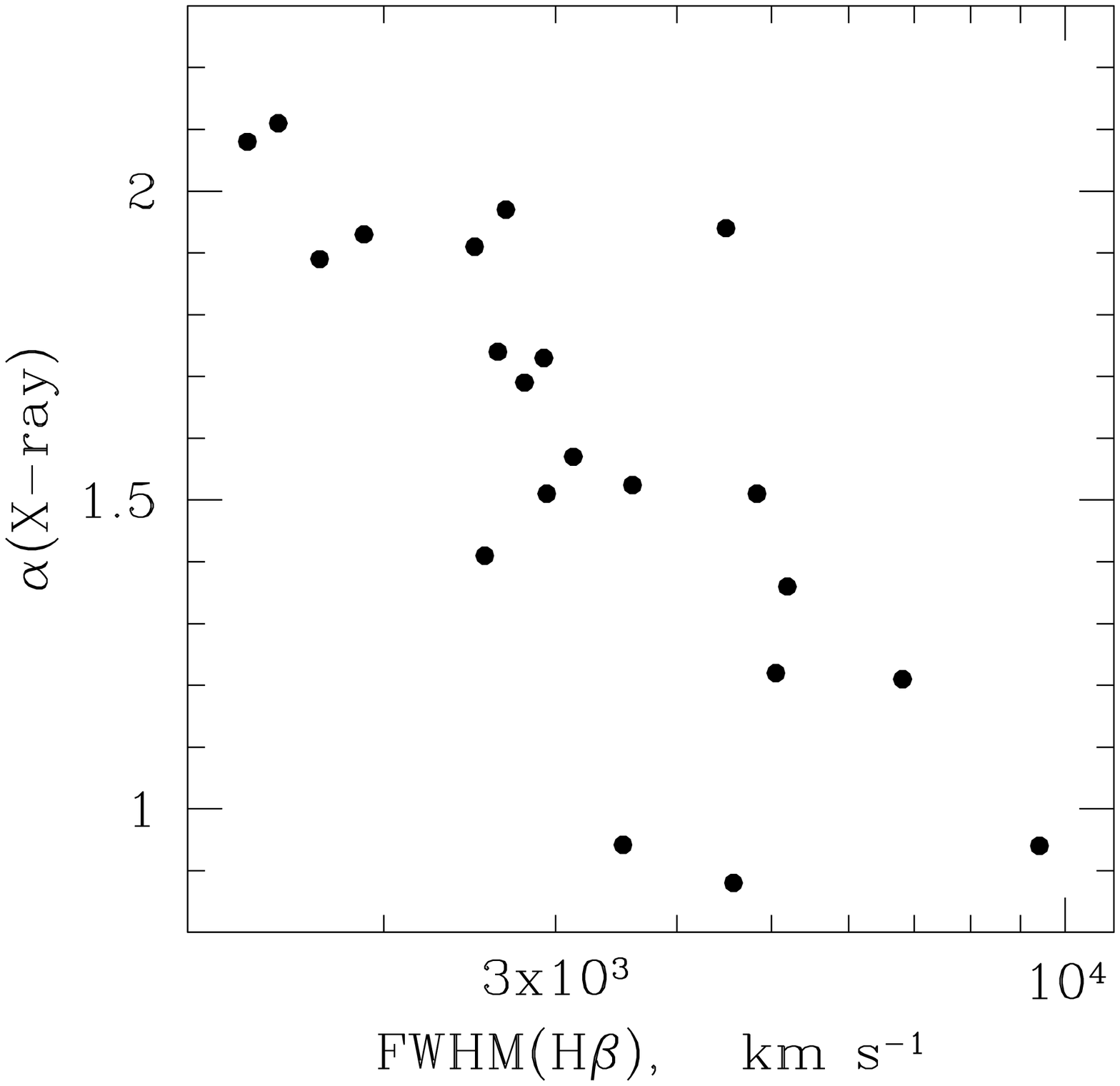}{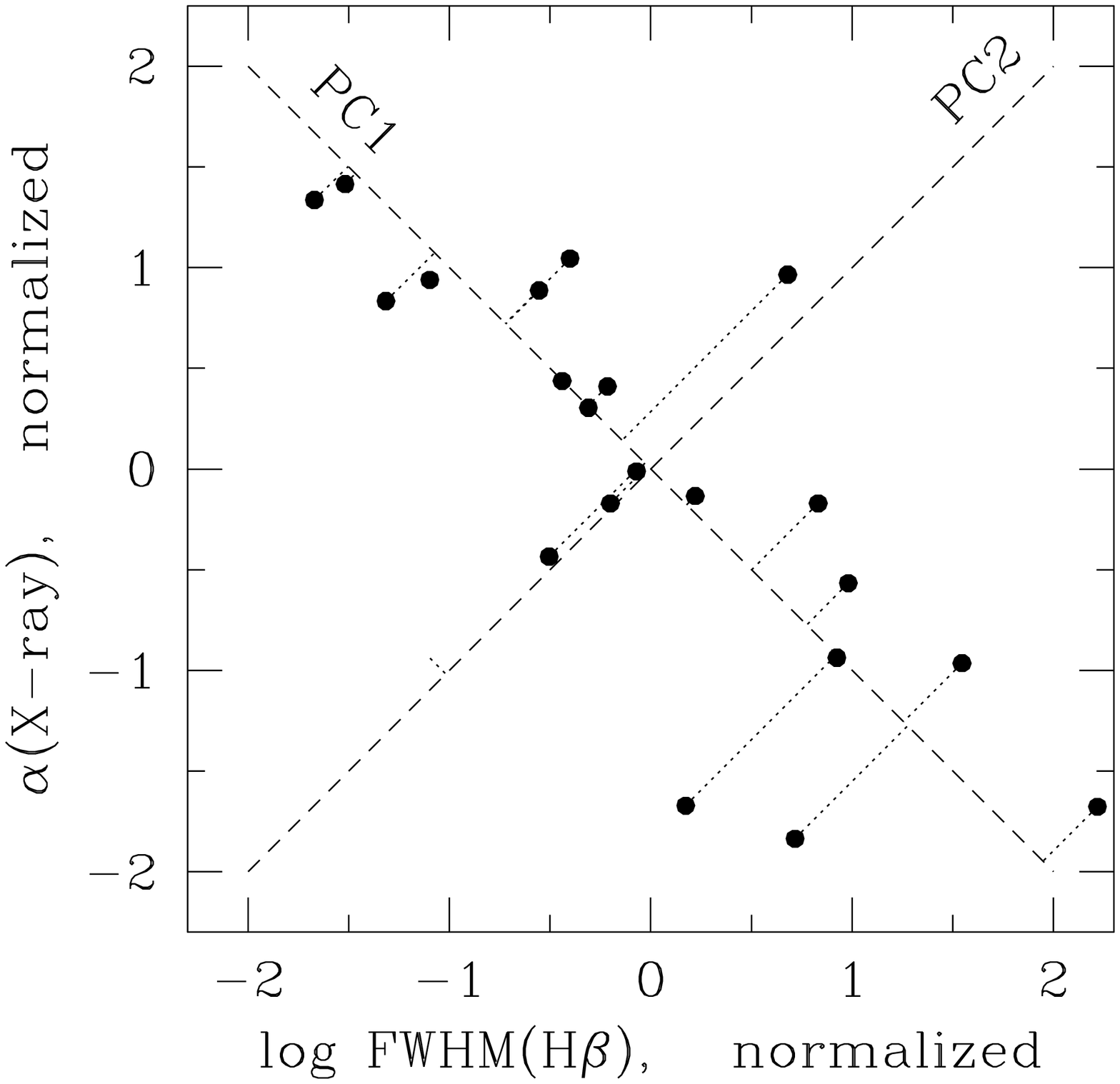}
\caption{An important optical correlation, soft X-ray spectral
index vs. width of the broad H$\beta$ emission line.
Left:  In natural units.
Right: In normalized units, with mean subtracted, then divided by
the standard deviation.  The dashed line shows the direction of
the first principal component (PC1), representing the maximum deviation of 
the cloud of
data points.  Dotted lines project the data points onto this direction.
PC1 represents the direction that minimizes the sums of the squares of
the lengths of the dotted lines.  The value (score) of PC1 for a given point
is the distance of the point from the origin, projected onto PC1.
Similarly the lengths of the dotted lines represent the values of PC2 for
each data point.
}
\end{figure}

We have succeeded in defining a new variable, a linear combination of
$\alpha_x$ and log FWHM H$\beta$, that accounts for most of the
variation within the sample (PC1).  The interpretation of this parameter is a
hotly contended topic (e.g., Pounds, Done \& Osborne 1995, Laor et al. 1997,
Brandt \& Boller 1998).  Is PC2 of
any significance?  The astronomer, with
knowledge of the measurement uncertainties, may have more hope of addressing this.
If the original variables had been uncorrelated, we could still define PC1 and PC2
mathematically, but we would be no better off as a result of the analysis.

%
%
%
%

\subsection{PCA with More Variables
}
\index{corr!eigen!opt}
\small
\begin{table}
\caption{Input Data} \label{tbl-1}
\begin{center}
\begin{tabular}{lcccccc}
\hline
\\
PG Name                
		       &  log
                       &  $\alpha_x$
                       &  log FWHM
                       &  FeII/
                       &  log EW
                       &  log FWHM\\
                        &  L$_{1216}$\tablenotemark{a}
                        &  
                        &  H$\beta$
                        &  H$\beta$
                        &  [OIII]
                        &  CIII]\\
\tableline
\\
0947$+$396 & 45.66   & 1.51  & 3.684    &   0.23  &  1.18 & 3.520 \\
0953$+$414 & 45.83   & 1.57  & 3.496    &   0.25  &  1.26 & 3.432 \\
1001$+$054 & 44.93   & \ldots & 3.241    &  0.82  &  0.85 & 3.424 \\
1114$+$445 & 44.99   & 0.88  & 3.660    &   0.20  &  1.23 & 3.654 \\
1115$+$407 & 45.41   & 1.89  & 3.236    &   0.54  &  0.78 & 3.403 \\
1116$+$215 & 46.00   & 1.73  & 3.465    &   0.47  &  1.00 & 3.446 \\
1202$+$281 & 44.77   & 1.22  & 3.703    &   0.29  &  1.56 & 3.434 \\
1216$+$069 & 46.03   & 1.36  & 3.715    &   0.20  &  1.00 & 3.514 \\
1226$+$023 & 46.74   & 0.94  & 3.547    &   0.57  &  0.70 & 3.477 \\
1309$+$355 & 45.55   & 1.51  & 3.468    &   0.28  &  1.28 & 3.406 \\
1322$+$659 & 45.42   & 1.69  & 3.446    &   0.59  &  0.90 & 3.351 \\
1352$+$183 & 45.34   & 1.52  & 3.556    &   0.46  &  1.00 & 3.548 \\
1402$+$261 & 45.74   & 1.93  & 3.281    &   1.23  &  0.30 & 3.229 \\
1411$+$442 & 44.93   & 1.97  & 3.427    &   0.49  &  1.18 & 3.275 \\
1415$+$451 & 45.08   & 1.74  & 3.418    &   1.25  &  0.30 & 3.434 \\
1425$+$267 & 45.72   & 0.94  & 3.974    &   0.11  &  1.56 & 3.666 \\
1427$+$480 & 45.54   & 1.41  & 3.405    &   0.36  &  1.76 & 3.300 \\
1440$+$356 & 45.23   & 2.08  & 3.161    &   1.19  &  1.00 & 3.192 \\
1444$+$407 & 45.92   & 1.91  & 3.394    &   1.45  &  0.30 & 3.479 \\
1512$+$370 & 46.04   & 1.21  & 3.833    &   0.16  &  1.76 & 3.546 \\
1543$+$489 & 46.02   & 2.11  & 3.193    &   0.85  &  0.00 & \ldots\\
1626$+$554 & 45.48   & 1.94  & 3.652    &   0.32  &  0.95 & 3.631 \\
\\
Number     & 22      & 21    & 22       & 22      & 22    & 21    \\ 
Mean       & 45.56   & 1.57  & 3.498    &   0.56  &  0.99 & 3.446 \\
Std dev'n  &  0.47   & 0.38  & 0.212    &   0.40  &  0.47 & 0.129

\end{tabular}
\end{center}
\tablenotetext{a}{Log of continuum luminosity at 1216\AA\ in units of erg s$^{-1}$ 
(H$_o = 50$ km s$^{-1}$ Mpc$^{-1}$, q$_o = 0.5$.) 
FWHM are in km s$^{-1}$; rest-frame equivalent widths (EW) are in \AA.
}
\end{table}

\normalsize
PCA achieves its real usefulness in multivariate problems.
We perform a PCA\footnote{Several widely available statistical 
packages include a task
for Principal Components Analyses (Statistical Package for the Social Sciences -- SSPS,
Statistical Analysis System -- SAS, Minitab -- Minitab Reference Manual 1992).}
on the small sample of 22 QSOs discussed by Wills et al. (1998a,b),
using a subset of the available measured properties shown in Tables 1 and 2.  
Unavoidably, there are missing data, so the number of objects available
depends on the variables chosen for the PCA.  Many
correlations are plotted, and the Pearson correlation coefficients tabulated, by
Wills et al. (this volume).
\small
\begin{table}
\caption{Input Data, continued} \label{tbl-1}
\begin{center}
\begin{tabular}{lcccccccc}
\hline
\\
PG Name                
		       &  log EW
		       &  log EW
                       &  CIV/
                       &  log EW
                       &  SiIII/
                       &  NV/
                       &  $\lambda$1400/\\
                        &  Ly$\alpha$
                        &  CIV
                        &  Ly$\alpha$
                        &  CIII]
                        &  CIII]
                        &  Ly$\alpha$
                        &  Ly$\alpha$\\
\tableline
\\
0947$+$396 &  2.08   & 1.78  &  0.45    & 1.24    & 0.306   &  0.179   &   0.143 \\
0953$+$414 &  2.19   & 1.78  &  0.40    & 1.24    & 0.164   &  0.189   &   0.093 \\
1001$+$054 &  2.25   & 1.76  &  0.40    & 1.43    & 0.443   &  0.462   &   0.174 \\
1114$+$445 &  2.27   & 1.85  &  0.42    & 1.48    & 0.222   &  0.175   &   0.092 \\
1115$+$407 &  1.90   & 1.51  &  0.33    & 1.14    & 0.385   &  0.228   &   0.134 \\
1116$+$215 &  2.14   & 1.71  &  0.34    & 1.20    & 0.440   &  0.254   &   0.126 \\
1202$+$281 &  2.72   & 2.41  &  0.69    & 1.87    & 0.164   &  0.154   &   0.098 \\
1216$+$069 &  2.12   & 1.95  &  0.54    & 1.20    & 0.037   &  0.121   &   0.056 \\
1226$+$023 &  1.64   & 1.44  &  0.45    & 1.00    & 0.280   &  0.174   &   0.018 \\
1309$+$355 &  2.01   & 1.68  &  0.41    & 1.15    & 0.303   &  0.131   &   0.064 \\
1322$+$659 &  2.19   & 1.85  &  0.41    & 1.30    & 0.291   &  0.135   &   0.097 \\
1352$+$183 &  2.14   & 1.80  &  0.41    & 1.29    & 0.357   &  0.203   &   0.116 \\
1402$+$261 &  1.91   & 1.59  &  0.39    & 1.09    & 0.568   &  0.227   &   0.161 \\
1411$+$442 &  \ldots & 1.88  &  \ldots  & 1,42    & 0.314   & \ldots   &   0.093 \\
1415$+$451 &  2.32   & 1.78  &  0.29    & 1.40    & 0.688   &  0.210   &   0.142 \\
1425$+$267 &  \ldots & 2.17  &  \ldots  & 1.43    & 0.398   & \ldots   &   0.055 \\
1427$+$480 &  2.03   & 1.82  &  0.49    & 1.21    & 0.265   &  0.126   &   0.117 \\
1440$+$356 &  2.14   & 1.54  &  0.21    & 1.05    & 0.747   &  0.141   &   0.092 \\
1444$+$407 &  1.99   & 1.34  &  0.21    & 1.06    & 0.809   &  0.335   &   0.164 \\
1512$+$370 &  2.02   & 2.05  &  0.75    & 1.28    & 0.228   &  0.182   &   0.050 \\
1543$+$489 &  1.93   & 1.60  &  0.44    & \ldots  & \ldots  &  0.398   &   0.174 \\
1626$+$554 &  2.14   & 1.80  &  0.39    & 1.36    & 0.197   &  0.217   &   0.118 \\
\\
Number     &  20     & 22    & 20       & 21      & 21      & 20       &  22  \\
Mean       &  2.11   & 1.78  & 0.421    & 1.279   & 0.362   &  0.212   &  0.108  \\
Std dev'n  &  0.21   & 0.24  & 0.131    & 0.194   & 0.199   &  0.091   &  0.043

\end{tabular}
\end{center}
\tablenotetext{}{ 
}
\end{table}

\normalsize
\index{corr!eigen!UV}
Tables 1 and 2 also present, for each variable, the number of data points, the mean and 
the standard deviation.
Notice the completely different units for
different measured parameters.  Clearly, from
our two dimensional example, one can see in Fig. 1 or 2 that the deviations from PC1,
hence PC1 itself, will depend on the units chosen (the weighting of the variables).
In order to weight the variables more or less equally, after subtracting the mean values,
we normalize by the
variance.  The choice of weights is a difficult issue, and depends on the user's
knowledge of the data, and preferences, as well as the use to which the results will be put.
The results of performing a PCA on these normalized variables are shown in Table 3.
Columns (2)--(6) show the first 5 out of a total of 13 principal components.  
The first row gives the variances (eigenvalues) of the data along the direction of the
corresponding principal component.  The sums of
all the variances add up to the sums of the variances of the input variables, in this case,
13.  By convention, the principal components are given in order of their contribution to
the total variance.  This is
given as `Proportion' in the second line, and the `Cumulative' proportion on the third line.
Thus, among the parameters we have chosen to use, the first principal component contributes
50\% of the
spectrum-to-spectrum variance, the second 22\%, the third, 12\%.  The first two principal
components together contribute 71\% of the variance, the first 3, 84\%, and the first 4,
nearly 90\%.

\small
\begin{table}
\caption{Results of Eigenanalysis -- The Principal Components\tablenotemark{a}} \label{tbl-2}
\begin{center}
\begin{tabular}{lrrrrr}
\tableline
\\
                             &   PC1    &  PC2    &  PC3   &  PC4  & PC5 \\
\tableline
\\
 Eigenvalue                & 6.4505  & 2.8157  & 1.5879  & 0.6257  & 0.5698    \\
 Proportion                &  0.496  &  0.217  &  0.122  &  0.048  &  0.044    \\
 Cumulative                &  0.496  &  0.713  &  0.835  &  0.883  &  0.927    \\
\\
 Variable                  &   PC1    &  PC2    &  PC3   &  PC4  & PC5 \\
\tableline
\\
log L$_{1216}$             &  0.053    &  0.535    & $-$0.123  & $-$0.029  & $-$0.405 \\
$\alpha_x$                 &  0.295    & $-$0.198  &  0.079    &  0.485    & $-$0.155 \\
FWHM H$\beta$              & $-$0.330  &  0.077    & $-$0.357  & $-$0.082  & $-$0.141 \\
Fe\,II/H$\beta$            &  0.341    & $-$0.140  &  0.003    & $-$0.487  & $-$0.212 \\
log EW $[$O\,III]          & $-$0.310  &  0.016    &  0.255    &  0.394    & $-$0.095 \\
log FWHM C\,III]           & $-$0.198  &  0.077    & $-$0.623  &  0.054    &  0.402 \\
log EW Ly$\alpha$          & $-$0.177  & $-$0.502  & $-$0.006  & $-$0.143  &  0.033 \\ 
log EW C\,IV               & $-$0.336  & $-$0.262  &  0.048    & $-$0.050  & $-$0.303 \\
C\,IV/Ly$\alpha$           & $-$0.342  &  0.062    &  0.025    & $-$0.074  & $-$0.584 \\
log EW C\,III]             & $-$0.262  & $-$0.413  & $-$0.124  & $-$0.176  & $-$0.008 \\
Si\,III]/C\,III]           &  0.342    & $-$0.149  & $-$0.018  & $-$0.311  & $-$0.116 \\
N\,V/Ly$\alpha$            &  0.231    & $-$0.050  & $-$0.573  &  0.107    & $-$0.288 \\
$\lambda$1400/Ly$\alpha$   &  0.223    & $-$0.351  & $-$0.225  &  0.441    & $-$0.216
\end{tabular}
\end{center}
\tablenotetext{a}{18 of 22 QSO spectra used; 4 cases contain missing values.
}
\end{table}

\normalsize
The columns of numbers for each principal component represent the weights assigned to each
input variable.  Thus PC1 $= 0.053\times x_1 + 0.295\times x_2 - 0.330 \times x_3 +$ 
\ldots, where $x_1$, $x_2$, and $x_3$ are the values of the normalized variables 
corresponding to log L$_{1216}$, $\alpha_x$, FWMH H$\beta$, etc.  By convention these 
weights are chosen so that
the sum of their squares = 1.  This arbitrarily fixes the scale of the new variable.  The sign
of the new variable is therefore arbitrary.

\subsection{Interpretation}

\index{corr!eigen!UV}
The first principal component is elongated with variance 6.5 times that of any individual
measurements, and accounts for about half the total variance.  This is therefore likely
to be highly significant.  If all measured, normalized quantities contributed equally
to PC1, they would all have weight 0.277 ($1/\sqrt{13}$ for 13 variables), but
each variable contributes more or less than this.  One way to test the significance of the
contribution of any one measured variable, is to perform the PCA without that variable, then
check the significance of the correlation between that variable and the scores of the new
principal component.  This procedure shows that all measured variables except L$_{1216}$, 
log FWHM CIII], and log EW Ly$\alpha$, correlate with PC1, but correlations involving  
NV/Ly$\alpha$ and $\lambda$1400/Ly$\alpha$ are not very strong.  PC2, accounting for 22\%
of the variance in this
dataset,  appears to link the EW Ly$\alpha$, EW CIV, and EW CIII] with L$_{1216}$, so
EW CIV and EW CIII] appear to contribute to both PC1 and PC2, but EW Ly$\alpha$ contributes
predominantly to PC2.  Is PC2 a significant component?  A similar correlation test
shows that individually the EWs do anti-correlate with L$_{1216}$, but this result depends
on the lowest EWs for the highest luminosity QSO PG1226+023 and the highest EWs for the low
luminosity QSO PG1202+281.  However L$_{1216}$ correlates significantly (Pearson's
ordinary correlation coefficient $= -0.77$)
with PC2 formed when L$_{1216}$ is excluded.  Thus there is a significant overall
correlation between EW and L$_{1216}$, although a larger sample is clearly needed
to investigate the individual EW correlations.
Another test may be to check correlations between observed measurements for those 
measurements
that contribute to only one significant principal component
 -- for example, C\,IV/Ly$\alpha$ vs.
Fe\,II/H$\beta$ (see Fig. and Table of Wills et al. in this volume).

As a rule-of-thumb, any principal component with variance greater than 1, should be 
considered
seriously.  It is also worth investigating any principal component with variance rather
greater than that of the remaining principal components.  In our example, this could mean
the first three principal components.

PCA is a linear analysis.  Tests should be performed to check on the linearity of the
principal components.  If a linear analysis is valid, plotting the scores of PC1 vs. PC2 
should show a normal distribution consistent
with no correlation between the two.  Mathematically, there cannot be a correlation,
but a non-random
distribution of points, or individual outlying points, may indicate non-linearity of the
relationships -- or some other problem with the uniformity of the data-set.  Outliers
could be rejected and the analysis repeated, or a transform of co-ordinates, for example
to logarithmic co-ordinates, may reduce the problem to a linear analysis.
A PCA performed using the ranks rather
than the actual (normalized) measurements may be more robust to both non-random
distributions and outliers.  (Compare the present results with those from the
analysis of the ranks, in Table 2 of the other PCA paper in this volume.)
These tests are an important tool for examining non-linearities in the data, and
for discovering individual unusual objects. 

\section{Some Examples from the Literature}

Increasing awareness of statistical methods has led to the
establishment of the Statistical Consulting Center For Astronomy at Penn State
University (Akritas et al. 1997, Feigelson et al. 1995, see also
http://www.stat.psu.edu/scca/ and www.astro.psu.edu/statcodes), and a series of
conference and other volumes devoted to statistics in astronomy (Murtagh \&
Heck 1987; Feigelson, Babu, \& Jogesh 1992), including PCA.

PCA is being increasingly applied in astrophysics.
Investigations of low and high redshift galaxies depend on their
classification (by morphology, photometry, kinematics, etc.), in terms of the
purely observational ``fundamental plane'', a subspace of the $p$-dimensional
parameter space (Djorgovski \& Davis 1987, see also an interesting PCA paper by M.
Han 1995).
The same area of astronomy has also extensively applied `neural network'
techniques for the automated classification of galaxy images, and more
(Odewahn 1998; Rawson, Bailey \& Francis 1996.)

An example similar to that presented here, using a subset of the parameters we consider,
but for a much larger sample, is provided in the paper by Boroson \& Green (1992).
Other examples of PCA analyses are given and discussed by
Whitney (1983a, b), and Murtagh and Heck (1987).  Some PCAs
have a larger number of variables than
input observables, $p$ $>$ $n$.  This results in a singular matrix and
therefore requires modifications to the techniques to solve the
eigenvector equations.  These techniques are discussed, for
example, by Wilkinson (1978), and mentioned by Mittaz, Penston, \& Snijders (1990).
This situation occurs in `Spectral PCA'.  The principles are
identical, but the number of variables is larger than the number
of QSO spectra.  Here the QSO spectra are divided into many
discrete bins, by wavelength or log (wavelength) (or velocity), and
the $p$ variables are the fluxes in these $p$ bins.  An excellent example
and discussion of interpretation is given by Francis et al. (1992)\footnote{
The spectral PCA code is available from the web:
http://msowww.anu.edu.au/\~\,pfrancis/
}.
For another example, see Wills, Brotherton, Wills \& Thompson (1997).
Spectral PCA also finds application to spectral time variability.  For
example, Mittaz et al. (1990) analyze the spectra
of NGC\,4151 at 59 epochs, binning each spectrum in wavelength space
(1375 bins).  A more recent example is given by T\"urler \& Courvoisier
(1998).

Recommended for further reading, is chapter 6 from Manly's `Multivariate
Statistical Methods' (1994), which gives a good brief discussion of the
method, with useful insights into interpretation.
A more rigorous mathematical treatment, together with discussion,
is given by {\it the}
great researcher and expositor of statistics  M. Kendall (Chapters 1 and
2 of `Multivariate Analysis'.)


\acknowledgments

We gratefully acknowledge many discussions with
M. S. Brotherton (Lawrence Livermore National Laboratory, Livermore, CA).
B.J.W. thanks Kris Berg, Dirk Grupe, and Sprout Berg for friendship and a
great deal of patience.
We also thank M. Cornell and R. Wilhelm who provide computer support in
the Astronomy Department of the University of Texas.
B.J.W. is supported by NASA through LTSA grant number NAG5-3431 and grant number
GO-06781 from the Space Telescope Science Institute, which is operated by
the Association of Universities for Research in Astronomy, Inc., under
NASA contract NAS5-26555.
We used MINITAB.

\end{document}